\documentstyle[aps,psfig,floats,epsf,prl]{revtex}

\begin{document}
\tighten
\draft

\title{Theory of valence transitions in Ytterbium and Europium intermetallics}
\author{V. Zlati\'c$^{\dagger}$ and J.K. Freericks$^*$}
\address{$^\dagger$ Institute of Physics, 10000 Zagreb, P.O. Box 304, Croatia}
\address{$^*$Department of Physics, Georgetown University, Washington, DC
20057, U.S.A.}
\date{\today}
\maketitle

\begin{abstract}
The exact solution of the multi-component Falicov-Kimball model 
in infinite-dimensions is presented and used to discuss a new fixed 
point of valence fluctuating intermetallics with Yb and Eu ions.
In these compounds, temperature, external magnetic field, pressure,
or chemical pressure induce a transition between a metallic state 
with the f-ions in a mixed-valent (non-magnetic) configuration and 
a semi-metallic state with the f-ions in an integral-valence 
(paramagnetic) configuration.
The zero-field transition occurs at the temperature $T_V$, while 
the zero-temperature transition  sets in at the critical field $H_c$.  
We present the thermodynamic and dynamic properties of the model 
for an arbitrary concentration of d- and f-electrons. 
For large U, we find a MI transition, triggered by the temperature 
or field-induced change in the f-occupancy. 
\end{abstract}

\section{Introduction}

Intermetallic compounds like YbInCu$_4$ 
\cite{felner.86,sarrao_review,ocko_Ag,ocko_Y,figueroa.98} 
and EuNi$_2$(Si$_{1-x}$Ge$_x$) \cite{levin.90,eu_japan}
exhibit valence-change transitions with dramatic effects 
on their thermodynamic and transport properties at
ambient pressure and low temperatures.  One characteristic behavior is
seen in the magnetic susceptibility which increases with a Curie-like or 
Curie-Weiss-like law from high temperatures up to the transition temperature
$T_V$ where it changes to a Pauli-like law with one to two orders of 
magnitude drop at the transition.  Similarly, a magnetic field, 
for $T<T_V$, induces a metamagnetic transition where
the magnetization increases rapidly in a field on the order of a few Tesla.
In the high-temperature phase the Yb ions are close to the 3+ (4f$^{13}$) 
configuration with one paramagnetic (J=7/2, L=3, S=1/2) f-hole, 
while the Eu ions are close to the 2+ (4f$^{7}$) configuration 
which has one paramagnetic (J=7/2, L=0)  f-electron. X-ray 
edge and Mossbauer spectroscopies indicate a valence change occurs 
at $T_V$ which is typically about 0.1 for YbInCu$_4$ and 0.6 for 
EuNi$_2$(Si$_{1-x}$Ge$_x$).  The volume change of the lattice at the
transition roughly agrees with the change in the size of the rare-earth 
ion assuming that the low-temperature phase of Yb ions contains a mixture 
of magnetic 3+ and non-magnetic 2+ (4f$^{14}$) configurations, and in the 
Eu compounds a mixture of the magnetic 2+ and non-magnetic 3+ (4f$^{6}$) states. 
The surprising feature is that despite the presence of well defined local 
moments there is 
no sign of the Kondo effect in the high temperature phase:
the electrical resistance is large and with a positive (temperature) slope, 
the thermoelectric power is small, the magnetoresistance is positive, 
and the Hall coefficient is large
\cite{sarrao_review,ocko_Ag,ocko_Y,figueroa.98}.
Hence, the high-temperature phase seems to be a semi-metal consisting 
of conduction electrons that scatter off of f-electron ions with 
well defined moments that behave nearly independently. 

As the temperature is reduced, the local moments become unstable, and the
system undergoes a transition to a low-temperature valence-fluctuating
metal that no longer possesses local moments. 
Magnetic and neutron scattering experiments\cite{severing.90} show that 
the loss of moment in the low-temperature phase is not due to the onset 
of any long-range order.  The low-temperature phase is  metallic with 
a low resistivity and a small Hall coefficient, but with a large peak 
in the thermopower. 
This loss of f-electron moment occurs not due to a complete transfer of the 
f-electron into the conduction band, but rather occurs due to the additional
formation of a hybridized valence-fluctuating phase with a quenched 
magnetic moment. But this valence-fluctuating phase is not an ordinary one,
because it is sensitive to external perturbations like a magnetic
field which can easily restore the local moments via a metamagnetic transition 
or hydrostatic (and chemical) pressure which stabilizes the f-electron
shell with a local moment.  Finally, optical conductivity measurements 
\cite{Garner.00} show a 
complete reconstruction of the many-body excitations as one passes 
through $T_V$, where a mid-infrared peak instantly appears in the 
low-temperature phase and a Drude peak sharpens and increases its spectral
weight.

A complete theoretical description of these systems poses quite a challenge.
The creation of a valence-fluctuating state in a conventional
periodic Anderson model is robust and insensitive to external perturbations
like temperature, magnetic field, or pressure. Transitions from the 
valence-fluctuating state to any other phase are all believed to be
continuous crossovers that are never first-order-like. On the other hand, the
stabilization of intermediate-valence states in the 
Falicov-Kimball\cite{falicov_kimball} (FK)
model is problematic in the limit as $T\rightarrow 0$, since the f-electrons
are usually completely filled or empty in the homogeneous ground state
phases. This is true even though the Falicov-Kimball model is known to
possess discontinuous valence-change transitions. A proper treatment of
these valence fluctuating systems requires combining the periodic Anderson
model with the Falicov-Kimball model, by introducing d-f Coulomb interactions
in the former or hybridization in the latter.  Solving such a model, and 
evaluating the phase diagram as a function of all of its parameters is
difficult, even with the advanced techniques of dynamical mean field theory.
Here we choose an alternate, and hence incomplete approach, which nevertheless
appears to capture the majority of the interesting behaviors of these
remarkable compounds.  We choose to describe these materials by a spin
one-half Falicov-Kimball model.  We will see below what properties can be
described by this model, and more importantly, we point out all of the
pitfalls that arise due to the neglect of the hybridization.  
We give reasonable arguments about how we expect hybridization 
to modify the behavior below.

The Falicov-Kimball Hamiltonian is
\begin{equation}
H=-\frac{t^*}{2\sqrt{d}}\sum_{\langle i,j\rangle\sigma}d^\dagger_{i\sigma}\
d_{j\sigma}+E_f\sum_{i\sigma}f^\dagger_{i\sigma}f_{i\sigma} +
U\sum_{i\sigma\sigma^\prime}f^\dagger_{i\sigma}f_{i\sigma}
c^{\dagger}_{i\sigma^\prime}c_{i\sigma^\prime},
\label{eq: ham}
\end{equation}
where $d^{\dagger}_{i\sigma}$ ($d_{i\sigma})$ is the electron creation 
(annihilation)
operator for an electron at site $i$ with spin $\sigma$, $E_f$ is the energy
level of the localized electrons (not to be confused with a Fermi
energy), $f$ labels the corresponding f-electron
creation and annihilation operators, and $U$ is
the interaction strength.  The total f-electron occupation is restricted to be
less than or equal to one at each lattice site, so that this system is
always strongly correlated ($U_{ff}\rightarrow\infty$). The hopping integral 
is scaled with the spatial dimension $d$ so as to have a finite 
result in the limit\cite{metzner_vollhardt} $d\rightarrow \infty$; 
we measure all energies in units of $t^*=1$. 
We work on a hypercubic lattice where the noninteracting density 
of states is a Gaussian $\rho(\epsilon)=\exp(-\epsilon^2)/\sqrt{\pi}$.
A chemical potential $\mu$ is used to maintain the total electron 
filling $n_{tot}=n_d+n_f$ at a constant value. 

The solution of the Falicov-Kimball model has been described before in
the literature\cite{brandt,freericks,zlatic} and won't be repeated here.  
Instead, we focus on the physical behavior of the exact solution.  
We present the results for the case where the total filling is 
$n_{tot}=1.5$ but the electrons can change from localized to itinerant 
(i.e., we fix the total electron concentration not the individual 
electron concentrations). 
We choose values of the parameters\cite{freericks,zlatic} where the system 
has a well defined crossover from a state at high temperature that
has large f-occupancy ($n_f \geq 0.36$ for $T\geq 0.2$), to a state at 
low temperature with no f-electrons (for definition of the transition 
temperature $T_V$, see below). To open a gap in the DOS at high temperatures we 
need a large f-f correlation and we choose $U=4$ (see below).  
To model the low-temperature transition, we need to tune the renormalized 
f-electron level to lie just above the chemical potential as 
$T\rightarrow 0$, and to describe YbInCu$_4$-like systems 
we choose several values of $E_f$. 
We find that the qualitative features of the results do not depend 
too strongly on the parameters in this regime (i.e., if we choose another total 
electron filling $1\le\rho_{\rm tot.}\le 2$, then we find similar results
when we vary $E_f$). As $T$ increases, 
the large entropy of the local moments stabilizes the f-electron number and
produces a large magnetic response of the system.  We choose the 
Coulomb interaction of an f-electron with a conduction electron to be such 
that the density of states splits into two bands\cite{freericks,zlatic}.
The shape and the relative weight of the two bands depends on the 
band filling and temperature but the peak-to-peak distance between the 
lower and the upper band is always about U.
The location of the f-level $E_f$, is the one parameter that we will 
adjust to control the high-temperature number of f-electrons 
(in the case of Eu) or f-holes (in the case of Yb).
Note, the single-particle excitations are strongly renormalized by the 
Coulomb interaction, and the bare value of $E_f$ has no physical meaning.
Since the FK model neglects the hybridization, we use the variation 
of $E_f$ to represent the change of the system due to the pressure or 
doping (chemical pressure). 
The gap in the density of states, which appears for $U\geq 1.5$, 
defines a characteristic energy scale $T^*$. The gap shrinks  
with the reduction of temperature but is fairly insensitive to the value 
of $E_f$ which, on the other hand, has a drastic effect on $T_V$.
For $U=4$ ($U=2$) we have $T^*\simeq 2$ ($T^*\simeq 0.4$) 
and $T^*\simeq 1.6$ ($T^*\simeq 0$, a pseudogap) at very high and 
low temperatures, respectively.  

In this paper we consider the system with $T^*\simeq 1.6$ ($U=4$) and 
discuss the temperature dependence of  
the spin susceptibility of f-electrons $\chi(T)$, the f-occupancy $n_f(T)$,  
the electrical resistance $\rho_{dc}(T)$, the thermoelectric power $S(T)$, 
and the thermal conductivity $\kappa(T)$, for five different values of $E_f$. 
We also discuss the frequency dependence of the optical 
conductivity $\sigma(\omega)$, the Raman response $B_{1g}(\omega)$, 
the magnetic field dependence of the f-electron magnetization 
$m_f(h,T)$ and the electrical resistance $R(h,T)$, 
at several characteristic temperatures. 
The finite-frequency results are presented for $E_f=-0.7$, and the 
finite-field data for $E_f=-0.6$, which are fairly typical values. 

\section{Results and discussion}

The first thing we show is a plot of $\chi(T)$ and $1/\chi(T)$, 
which is a Curie-like law, but with a temperature-dependent 
concentration of spins $\chi\propto n_f(T)/T$.
One can see from panel (a), in Figure 1 that as $E_f$ is made more negative, 
the peak in the susceptibility moves to lower $T$ and the curves sharpen.  
We define the valence transition temperature $T_V$ to be the temperature
below the peak in the susceptibility where $\chi(T_V)=\chi(T_{\rm peak})/2$,
i.e., it is at the temperature corresponding to half the peak height of the
susceptibility.  This definition provides a well-defined benchmark for
the value of $T_V$ that is similar to choosing the inflection point of the
susceptibility curves. All of the transitions we see here are
actually smooth crossovers centered around $T_V\ll T^*$.  
We believe there is a region of parameters where the transition 
can be made discontinuous at low $T$, but the iterative
algorithm normally used to solve this problem becomes unstable in the
region close to such first-order transitions, and alternate methods need
to be used to solve the problem.  In panel (b) we plot the inverse 
susceptibility which shows the characteristic Curie or Curie-Weiss law at 
moderate temperatures (as $E_f$ increases, the intercept of the Curie-Weiss
law becomes more negative).  Note that in our case there never is a low
temperature turnover of the inverse susceptibility, since we plot only the
f-electron susceptibility here, which vanishes as $n_f\rightarrow 0$.
In YbInCu$_4$ and EuNi$_2$(Si$_{1-x}$Ge$_x$)-like systems the 
ground state is a valence fluctuator with an enhanced Pauli 
susceptibility, such that the Curie plot shows a characteristic low-$T$ maximum. 

\begin{figure}[htfb]
                          \label{fig: chi}
\epsfxsize=3.5in
\centerline{\epsffile{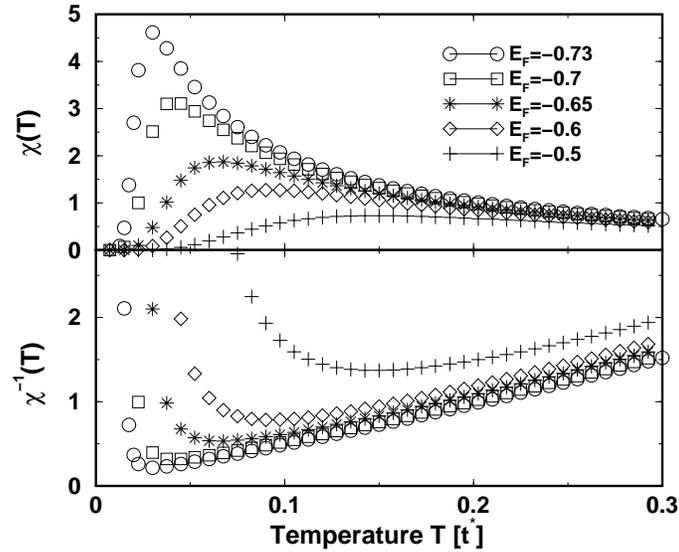}}
\caption{f-electron magnetic susceptibility (a) and inverse susceptibility
(b) for the Falicov-Kimball model with $n_{\rm total}=1.5$ and $U=4$.  The
symbols denote different values of $E_f$.  These curves look remarkably like 
the experimental curves at high temperature (and through the transition).}
\end{figure}

\begin{figure}[htfb]
                          \label{fig: n_f}
\epsfxsize=3.5in
\centerline{\epsffile{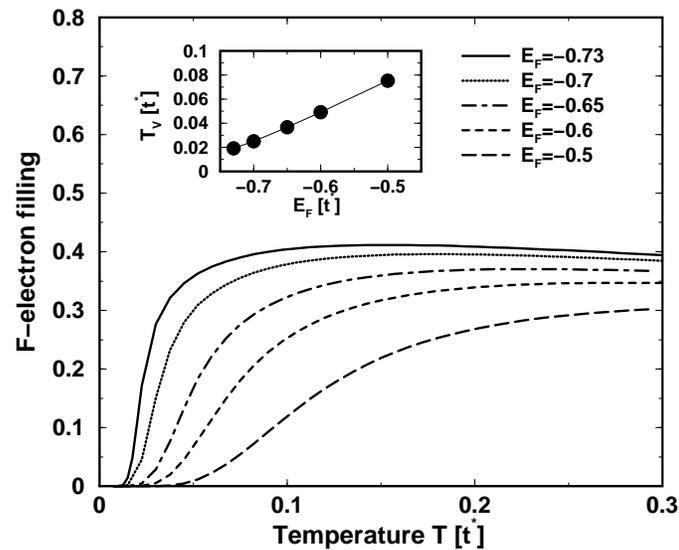}}
\caption{
f-electron filling versus temperature for the Falicov-Kimball model 
for the same parameters as in Fig.\ \ref{fig: chi}. 
The inset shows the valence transition temperature $T_V$ 
versus the parameter $E_f$.
}
\end{figure}
In Figure 2 we plot $n_f(T)$, which is relatively constant over 
a large temperature range and then suddenly drops toward zero at the 
transition.  As $E_f$ is made more negative, and the high-temperature 
f-filling is increased, the transition temperature
$T_V$ is reduced and the $n_f(T)$-curves 
sharpen. Note, the magnetic susceptibility, the particle number and 
the charge susceptibility, defined as $\chi_{c}=\partial n_f/\partial\mu$,  
reveal a characteristic scale $T_V$ but do not provide any 
hint on the magnitude of $T^*$.  
The inset shows the correlation between $T_V$ and $E_f$, 
which is our analog of experimental plots of $T_V$ versus doping
(for chemical pressure) or versus pressure.
The behavior seen here, where applying pressure makes the $E_f$ 
less negative and reduces the number of f-electrons, 
is similar to what would occur in experiment on Eu compounds; 
$T_V$ increases along with a smoothing out of the susceptibility. 
In Yb-based systems, pressure has the opposite effect\cite{sarrao_review} 
on the properties of the system, since it increases the number of f-holes.  
As mentioned already, and discussed in detail in 
Refs.~\onlinecite{freericks,zlatic}, because of the FK interaction 
the temperature- or field-induced change in $n_f$ leads to the MI transition 
in the conduction band, and is accompanied by substantial changes 
in the density of states. 
In the low-temperature phase, which is found for $T<T_V$, 
the weight of the upper conduction band is very small and 
the chemical potential is within the lower conduction band. 
In the high-temperature phase, which sets in for $T>T_V$, 
the spectral weight of the lower and upper band 
are about the same, and the chemical potential lies within the gap. 
Contrary to $T_V$, the characteristic temperature $T^*$ is only 
weakly dependent on $E_f$. 

\begin{figure}[htfb]
                         \label{fig: transport}
\epsfxsize=4.5in
\centerline{\epsffile{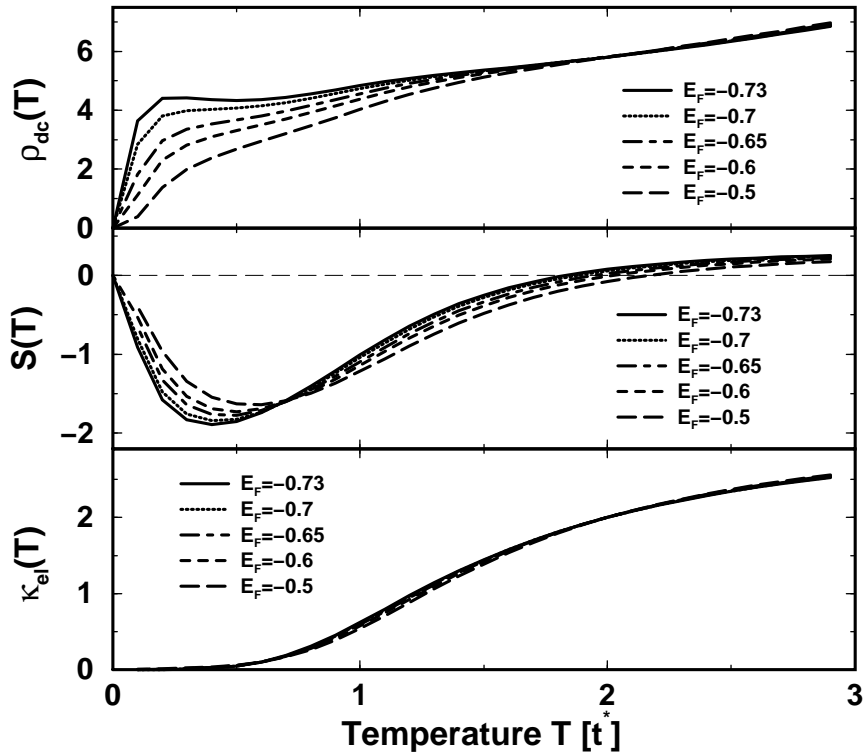}}
\caption{
DC conductivity (a), thermoelectric power (b), and 
thermal conductivity (c) 
for the spin-one-half Falicov-Kimball model 
as a function of temperature.  
The unrenormalized f-level is varied here to represent 
different doping or pressures. The high temperature 
phase is a poor metal with a moderate thermopower.
At low temperatures, a normal metallic state develops.
The thermal conductivity does not depend strongly 
on $E_f$ in this parameter regime. }
\end{figure}
The DC transport properties in zero magnetic field, 
obtained by an exact evaluation of the current-current, 
current-heat current, and heat current-heat current 
correlation functions, are summarized in Figure 3.  
We show $\rho_{dc}(T)$, $S(T)$, and $\kappa(T)$, using the 
normalization that is usual for infinite dimensional 
lattices\cite{pruschke.95}. The thermopower is plotted for 
the hole-like Yb case; for the Eu compounds, the sign would have 
to be changed. 
The high temperature phase is a poor metal with a minimum in $S(T)$ 
at about $T^*/3$ and small hump in $\rho_{dc}(T)$ at about $T^*/2$ (do not
confuse the hump at $T\approx 1$ with the sharp low-$T$ decrease in $\rho_{dc}$
that occurs near $T_V$).  Note, the thermopower 
does not exhibit any additional features around $T_V$. 
The minimum of $S(T)$ decreases from $T_{min}=0.6$ to $T_{min}=0.4$,  
as we change from $E_f=-0.5$ ($T_V=0.07$) to $E_f=-0.73$ ($T_V=0.03$), 
i.e. the relevant temperature scale for the thermoelectric power 
is set by $T^*$ and not by $T_V$. 
In the $U=2$ case, which has the high-temperature gap $T^*\simeq 0.4$, 
the minimum of $S(T)$ shifts from $T_{min}\simeq 0.4$ for $E_f=-0.2$ 
($T_V=0.25$) to $T_{min}\simeq 0.2$ for $E_f=-0.6$ ($T_V < 0.007$). 
The electronic part of the thermal conductivity shows a featureless 
decrease as temperature is reduced and no $E_f$-dependence. 
Thus, as long as $T_V\leq T^*$, the temperature scales revealed by the 
thermoelectric data and the thermodynamic data appear to be fundamentally 
different. 

The temperature dependence of various correlation functions discussed 
above is remarkably similar to what one observes in YbInCu$_4$ and 
EuNi$_2$(Si$_{1-x}$Ge$_x$)-like systems for $T\geq T_V$. 
The thermoelectric data obtained recently for Yb$_{1-x}$Y$_x$InCu$_4$ and 
YbAg$_x$In$_{1-x}$Cu$_4$ for various concentrations of Y and Ag 
dopants\cite{ocko_Ag,ocko_Y} can be explained by the FK model, 
if we take the initial parameters for YbInCu$_4$ such that 
$T_V \simeq  T^*\simeq T_{min}$, and assume that yttrium doping 
makes $E_f$ more negative, while silver doping does the opposite. 
Note, the thermoelectric power of the low-temperature valence fluctuating 
phase is much larger than of the high-temperature semimetallic 
phase\cite{ocko_Ag,ocko_Y}, i.e. the heating of real systems 
across $T_V$ always leads to a smaller $S(T)$.
The Y-doping pushes $T_V$ (as defined by the susceptibility anomaly) 
to lower temperatures but does not change the position of the 
high-temperature minimum of $S(T)$, since $T_{min}\simeq T^* > T_V$, 
and this does not depend on $x$ or $E_f$.
The thermoelectric power of Yb$_{1-x}$Y$_x$InCu$_4$ exhibits two minima: 
a deep one at the temperature of the valence transition and 
shallow one at the temperature of the pseudogap. The first one at $T_V$  
is doping-dependent, and the second one is doping-independent.
In YbAg$_x$In$_{1-x}$Cu$_4$ one finds the valence transition for $x\leq 0.3$, 
and the silver doping enhances $T_V$, i.e.  $T_V(x) > T^*$. 
The slope of $S(T)$ in the valence-fluctuating phase 
is characterized by the Kondo scale $T_{K} > T_V$, and as the temperature 
increases up to $T_V(x)$, $S(T)$ can reach large values. 
Above $T_V$, $S(T)$ drops to values typical of the 
high-temperature phase but since $T>T_V > T^*$, 
there is no high-temperature minimum of $S(T)$. 
Thus, as long as $T_{K} > T_V > T^*$,  the thermopower has a 
single minimum at a temperature that coincides with the susceptibility 
maximum, and is strongly doping (or $E_f$) dependent. 
However, Ag doping not only enhances $T_V$ but also reduces $T_{K}$, 
and for $x\geq 0.3$ the YbAg$_x$In$_{1-x}$Cu$_4$ becomes a heavy fermion 
with $T_{K} < T_V(x)$. For $T_{K}< T < T_V$, the Kondo effect stabilizes  
the paramagnetic state before the valence transition takes place and  
the thermoelectric power exhibits the usual Kondo minimum at about 
$T_{K}/2$, that is not much changed by further doping.  
The effect of hydrostatic pressure on these compounds\cite{sarrao_review}
follows from similar considerations, if we assume that pressure shifts 
$E_f$ and reduces (increases) the average number of f-electrons (f-holes). 

\begin{figure}[htfb]
                         \label{fig: optical}
\epsfxsize=4.5in
\centerline{\epsffile{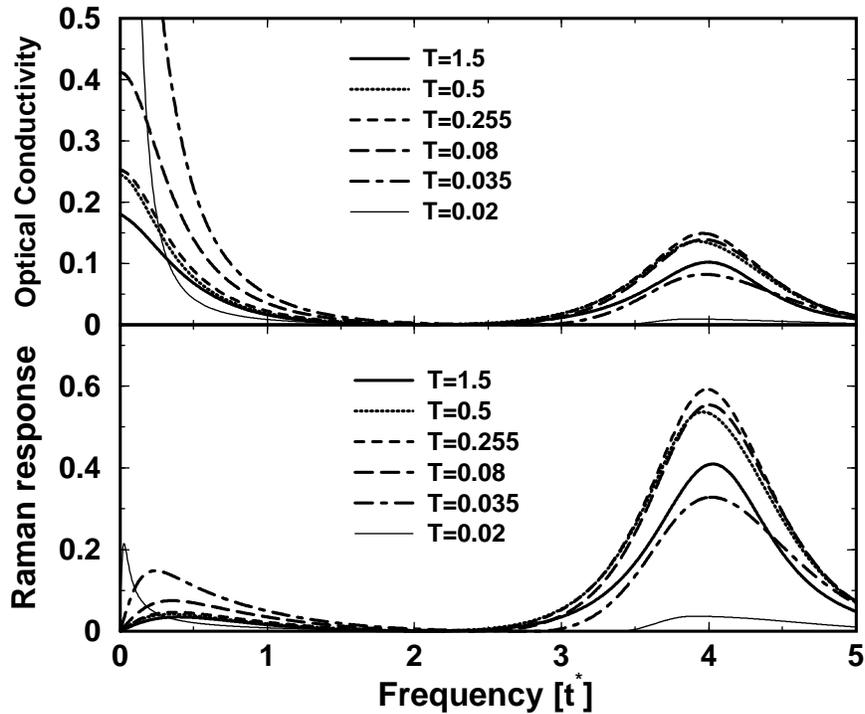}}
\caption{Optical conductivity (a) and $B_{\rm 1g}$ Raman response (b) of the
spin-one-half Falicov-Kimball model with $E_f=-0.7$ ($T_V=0.03$).  
Note how at high temperatures there is a separation 
of many-body excitations into
a low-energy Drude-like peak and a higher-energy charge-transfer peak.
As the temperature is lowered, spectral weight moves between these peaks,
in an expected way for a sharpening of the ``Fermi surface'', until one
reaches $T_V$ where the spectral weight changes dramatically, with the
charge-transfer peak losing weight and the Drude peak being strongly
enhanced.  This last behavior is an artifact of not being able to describe
the mixed-valence regime properly. }
\end{figure}
The transport properties of EuNi$_2$(Si$_{1-x}$Ge$_x$)$_2$ compounds, 
which show\cite{levin.90,eu_japan} (for $0.7\leq x \leq 0.8$) a valence 
transition at $T_V(x)\geq 85$ K, can also be explained by the FK model,  
but with f-electrons rather than holes.  
The reduction of the Ge concentration shrinks the lattice, reduces the 
average value of f-electrons, increases $T_V$, 
and makes the susceptibility transition less sharp. In this electron-like 
system, the thermoelectric power is positive\cite{levin.90}. 
Since the maximum of $S(T)$ is concentration dependent and 
$T_{max}\simeq T_V$, we would classify the high-temperature phase of 
EuNi$_2$(Si$_{1-x}$Ge$_x$)$_2$ for $0.7\leq x \leq 0.8$, as a small-gap 
FK state, with $T^*\leq T_V$.

In Figure 4, we plot the optical conductivity\cite{pruschke.95}
$\sigma(\omega)$ and the $B_{\rm 1g}(\omega)$ 
Raman response\cite{raman_us}
for the case $E_f=-0.7$ at a number of different temperatures.
In the high temperature phase, the optical conductivity has the characteristic 
poor metal behavior, with a low-energy Drude like feature and a large
weight charge-transfer peak around $\omega\simeq U$. 
As $T$ is lowered, the scattering processes become sharpened, 
and both peaks grow in temperature.  The charge-transfer peak is
frozen in at around $T=0.5$, but the Drude feature continues to increase.
As $T_V$ is approached, we see a transfer of weight from the charge-transfer
peak to the Drude peak, which then becomes complete as $T$ is lowered well
below $T_V$.  In this lowest temperature region, the results are not
representative of real materials, as we expect there to still be a 
charge-transfer peak since the f-electrons have not gone into 
the conduction band, they have only hybridized with them.  
We also expect midinfrared excitations to appear, because the 
hybridization with the f-electrons will allow them
to participate in the optical scattering, and the broadened f-spectral
function should have weight in the midinfrared region.  These latter
features are seen in experiment on the Yb compounds.  
The Raman response behaves similarly, with a sharpening of the 
charge-transfer peak and the growth of a lower Fermi-liquid peak 
as $T$ is lowered to $T_V$, where the spectral weight shifts 
become too large. Thus, the optical experiments can be used to 
estimate the value of U.    
We expect one would also see the turn on of the midinfrared response 
in the Raman scattering as $T$ is lowered below $T_V$
for the exact same reason it is seen in the optical conductivity.
If we estimate the f-d correlation from the high-frequency peak 
in the optical conductivity and use the susceptibility results 
to define $T_V$, we obtain from Figs.~4 and 2 
the ratio $U/k_B T_V=4/0.03=133$.  
If we use model with $U=2$ and $E_f=-0.6$, the ratio is $2/0.007=285$. 
The optical data for YbInCu$_4$ crystals, with $T_V\simeq 42$ K,  
exhibit the peak at 12000 cm$^{-1}$\cite{Garner.00}, which gives 
the experimental ratio $U/T_V=12000\;{\rm cm}^{-1}/42\;{\rm K}\simeq 375$. 

\begin{figure}[htfb]
                        \label{fig:field}
\epsfxsize=4.5in
\centerline{\epsffile{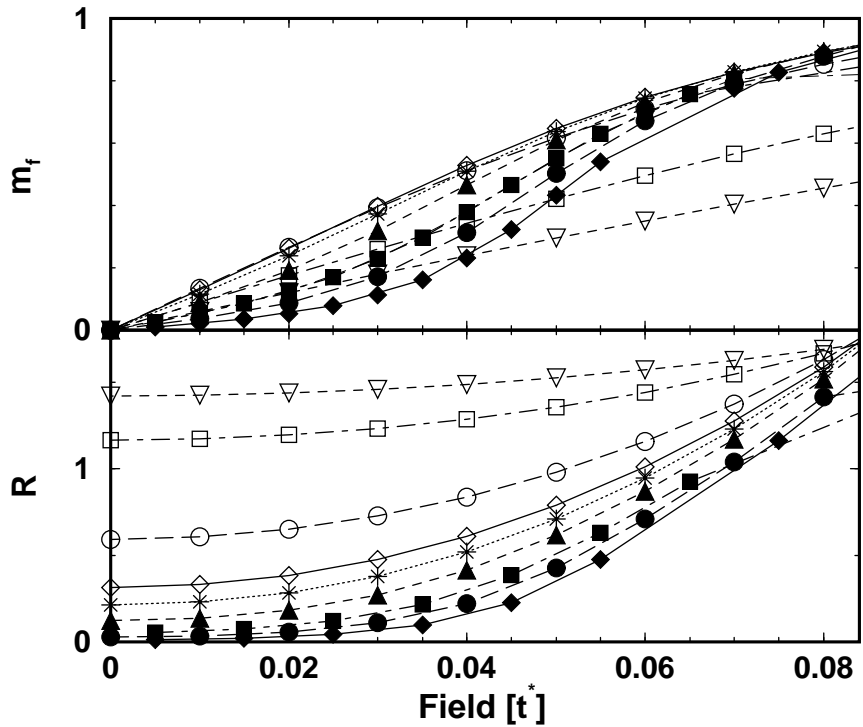}}
\caption{Magnetic field dependence 
of the f-electron magnetization (a) 
and the electrical resistance (b) 
for the spin-one-half Falicov-Kimball model 
with $n_{\rm total}=1.5$, $U=4$, $E_f=-0.6$ ($T_V=0.05$) 
The filled symbols connected by the full, long-dashed, dashed-dotted, dashed 
and dotted line correspond to temperatures $0.03 t^*$, $0.035 t^*$,
$0.04 t^*$, $0.05 t^*$, and $0.06 t^*$, respectively. 
The open symbols connected by the full, long-dashed, dashed-dotted, and 
dashed line correspond to temperatures $0.07 t^*$, $0.1 t^*$, 
 $0.2 t^*$, and  $0.3 t^*$, respectively.}
\end{figure}
The field dependence of $m_f(h,T)$ and $R(h,T)$ is shown in Figure 5 
for $U=4$, $E_f=-0.6$ ($T_V\simeq 0.05$). Each curve corresponds 
to a different temperature, as indicated in the caption. 
For low temperatures and fields, $m_f(h,T)$ and $R(h,T)$ behave typically 
of a normal Fermi liquid. At high temperatures and high fields, 
however, $m_f(h,T)$ behaves as expected of a non-interacting 
localized  moment, while $R(h,T)$ indicates a poor metal or a semiconductor.
Figure 5 shows that the low-temperature metallic state is 
destroyed by large enough magnetic field. At a critical field $H_c$ 
we find a metamagnetic transition in the f-subsystem and a metal-insulator 
transition in the conduction band. 
The H-T  phase boundary between the non-magnetic (metallic) phase and 
the paramagnetic (semimetallic) phase can be approximated by the 
expression $H_c(T)/H_c^0=\sqrt{1-(2T/T_V)^2}$, where the critical field 
$H_c(T)$ is estimated from the midpoint of the magnetization curves 
shown in Fig.\ 5, and  $H_c^0$ is the critical field at $T=0$. 
The ratio of the zero-temperature Zeeman energy at $H_c^0$ and the 
zero-field thermal energy at $T_V$ is $\mu_B H_c^0/k_B T_V\simeq 2$, 
where the gyromagnetic factor is $g=1$, since we are interested  
in the Yb or Eu compounds which have the f-electrons in their lowest 
spin-orbit state. Thus, the high-field and the low-field data  
show the same characteristic energy scale. 

\section{Conclusions}

We discuss the properties of YbInCu$_4$ and EuNi$_2$(Si$_{1-x}$Ge$_x$)-like 
compounds using the exact solution the Falicov-Kimball model in infinite 
dimensions.
In addition to the usual FK interaction, we include an infinitely
large repulsion between f-particles of the opposite spin, which leads
to $n_f\leq 1$ ($n_f$ describes electrons for Eu and holes for Yb).
We consider the total filling $n_{tot}=1.5$, and choose the parameters
such that the ground state has a large d-electron spectral weight around
the chemical potential $\mu$ and a negligible f-occupancy.
The mass of f-particles is assumed to be infinite but
the f-level has a finite width because of the FK coupling
to the density fluctuations in the conduction band.
(Not surprising, this feature is similar to what one finds in the X-ray edge
problem, defined by a single impurity version of the FK Hamiltonian.)
The spectral functions of the f- and d-states are gapped and the spectral
weight is strongly temperature- and field-dependent. At low temperatures,
there are many more conduction states below than above the gap, while
for the f-states just the opposite is true. At high temperatures,
the chemical potential is in the gap, and the weight of the occupied
and unoccupied the d- and f-states are about the same.

The FK model predicts a valence transition at $T_V$, a metamagnetic
transition for the f-electron magnetization, and a metal-insulator
transition in the conduction band, such that the high-temperature,
high-field transport is dominated by a pseudogap or a gap $\sim T^*$.
The paramagnetic, semimetallic phase sets in because of thermal
population of highly-degenerate f-states, which can be understood as an
entropy driven transition.
The proximity of the renormalized f-state to the chemical potential
explains the sensitivity of the transition to an external field,
doping or pressure. By tuning the position of the f-level, the
transition can be pushed to low temperatures, where it becomes very
sharp. 

For $T > T_V$, the number of f-particles is large and only weakly
temperature dependent. The zero-field susceptibility is given by the
Curie law, $\chi(T) \simeq n_f(T)/T$, rather than the Curie-Weiss law,
i.e. the Curie plot gives only a small Curie-Weiss temperature and a
``Curie constant" that depends on the thermal occupation of the excited
f-states. Note, the presence of the local moment in the high-temperature 
phase is not accompanied by the usual Kondo anomalies in the 
electronic transport, since the model has no f-d mixing.
The calculated electrical resistance is large with a positive (temperature)
slope, the magnetoresistance is positive, and the thermoelectric power
is small with a peak at about $T\simeq T^*$.
The optical conductivity has a pronounced peak at $\omega \simeq U$ and
a suppressed Drude peak, as expected of a strongly correlated semimetal.
Most of these features are observed in the high-temperature, high-field  
phase of YbInCu$_4$ and EuNi$_2$(Si$_{1-x}$Ge$_x$)-like compounds and they
seem to characterize a new fixed point of strongly correlated electrons.

The low-temperature, low-field properties of the FK model correspond to
a metal with a Pauli susceptibility, usual DC transport with a small
thermoelectric power, and a  large Drude peak in the optical conductivity.
However, the model neglects the d-f hybridization and its 
low-temperature metallic phase without the f-particles, 
cannot describe the low-temperature state of intermetallic
compounds with Eu and Yb ions.
The physical ground state is a valence fluctuator with a large average
f-occupancy, an enhanced Pauli susceptibility, huge thermoelectric
power, and an additional midinfrared peak in the optical conductivity.
All these features follow from the periodic Anderson model with an
infinite f-f correlation and a d-f hybridization.
In that model, the proximity of the d and f spectral weight to the chemical
potential leads to d-f inter-band transitions and a mid-infrared peak
in $\sigma(\omega)$. It appears that the low-temperature phase of
the above mentioned compounds can be explained by the PAM,
and the f-d correlation can be neglected.

On the other hand, the semimetallic high-temperature phase of
YbInCu$_4$ and EuNi$_2$(Si$_{1-x}$Ge$_x$) is difficult to describe
in terms of the usual solutions of the PAM, which can lead to a
fluctuating local f-moment but always keeps the d-electrons metallic.
The absence of the Kondo effect in the electrical resistance, the
magnetoresistance and the thermoelectric power data is difficult to
reconcile with the susceptibility data which is typical of local moments
with a small Curie-Weiss temperature. Such a behavior is not easy
to explain in terms of conduction electrons that exchange scatter
on the localized f-moments. As long as the f-state overlaps the
continuum of  metallic states, the quantum mixing leads to the Kondo
effect, and that is not seen in the high-temperature phase of
YbInCu$_4$ and EuNi$_2$(Si$_{1-x}$Ge$_x$)-like compounds.
Also, it is not clear that the PAM could provide a sharp transition
from a metallic (mixed valence, non-magnetic) state to a semimetallic
(integral valence, paramagnetic) state, as observed experimentally.
In our approach, we argue that the high-temperature width and the weight 
of the f-spectral function below the gap is very large, such that 
the additional broadening due to the f-d mixing can be neglected. 
In that sense, the FK model can be considered as an effective
model for the high-temperature phase of a generalized periodic Anderson
model which, in addition to the usual f-f correlations and
f-d hybridization, also has the f-d correlation.
It would be interesting to study the low temperature fixed point of such a
generalized periodic Anderson model with the Falicov-Kimball term, to
see if it reduces to the usual valence fluctuating fixed point of the
periodic Anderson model.

\acknowledgments

We acknowledge support of the National Science Foundation under grant
DMR-9973225.  We also acknowledge useful discussions with S.L. Cooper, 
G. Czycholl, C. Geibel, M.O\v cko and J. Sarrao.

\end{document}